# Sensitivity Assessment of Multi-Criteria Decision-Making Methods in Chemical Engineering Optimization Applications

**Seyed Reza Nabavi [a], Zhiyuan Wang [b], Gade Pandu Rangaiah [c, d, \*]**

[a] Department of Applied Chemistry, Faculty of Chemistry, University of Mazandaran, Babolsar, Iran

[b] Department of Computer Science, DigiPen Institute of Technology Singapore, Singapore 139660, Singapore

[c] Department of Chemical and Biomolecular Engineering, National University of Singapore, Singapore 117585, Singapore

[d] School of Chemical Engineering, Vellore Institute of Technology, Vellore 632014, India

\* Corresponding author: Gade Pandu Rangaiah, Email: chegpr@nus.edu.sg

**Abstract**

This chapter assesses the sensitivity of multi-criteria decision-making (MCDM) methods to modifications within the decision or objective matrix (DOM) in the context of chemical engineering optimization applications. Employing eight common or recent MCDM methods and three weighting methods, this study evaluates the impact of three specific DOM alterations: linear transformation of an objective (LTO), reciprocal objective reformulation (ROR), and the removal of alternatives (RA). Our comprehensive analysis reveals that the weights generated by entropy method are more sensitive to the examined modifications compared to the criteria importance through intercriteria correlation (CRITIC) and standard deviation (StDev) methods. ROR is found to have the largest effect on the ranking of alternatives. Moreover, certain methods, gray relational analysis (GRA) without any weights, multi-attributive border approximation area comparison (MABAC), combinative distance-based assessment (CODAS), and simple additive weighting (SAW) with entropy or CRITIC weights, and CODAS, SAW, and technique for order of preference by similarity to ideal solution (TOPSIS) with StDev weight are more robust to DOM modifications. This investigation not only corroborates the findings from the previous study, but also offers insights into the stability and reliability of MCDM methods in the context of chemical engineering.

# 1. Introduction

Multi-criteria decision-making (MCDM) is a powerful tool to assist decision-makers navigate complex scenarios, where they must select one of the alternatives available, considering multiple criteria simultaneously. These criteria often conflict with one another, adding to the decision-making challenge. In recent years, MCDM has garnered considerable interest across a diverse range of fields, such as logistics (Görçün et al., 2023; Miškić et al., 2023), healthcare (Chakraborty et al., 2023; Ali et al., 2024), finance (Wang et al., 2023a; Elma et al., 2024), sustainable agriculture (Cicciù et al., 2022; Kumar & Pant, 2023), product design (Chen et al., 2020; Wang et al., 2023b), artificial intelligence (Freire et al., 2021; Alshahrani et al., 2024), and chemical engineering (AlNouss et al., 2024; Bhojane et al., 2023; Nabavi et al., 2023a; Wang et al., 2023c). As depicted in Figure 1, MCDM is oftentimes viewed as the direct subsequent step following multi-objective optimization (MOO) (Rangaiah et al., 2020). In applications involving multiple conflicting objectives, MOO produces a group of Pareto-optimal solutions (also referred to as non-dominated solutions). However, implementing all these solutions is impractical in real-world applications, thus necessitating the selection of only one solution using MCDM. In the context of MCDM and throughout this chapter, the term "criteria" is used interchangeably with "objectives", and "alternatives" with "solutions".

Previous works by Pamučar & Ćirović (2015), Mufazzal and Muzakkir (2018), Wang et al. (2021), Shih (2022), and Dehshiri & Firoozabadi (2023) showcased that MCDM methods struggle with maintaining consistency with respect to linearly transformed changes of measurement unit (which commonly occurs in chemical engineering), equivalent objective formulation (e.g., in a thermal cracking process, maximization of ethylene selectivity can also be expressed as the minimization of 1/selectivity or 1 – selectivity) and/or addition/removal of alternatives (that could happen in case MOO did not find all Pareto-optimal solutions in the first run, and subsequently, additional solutions were added in later runs). The sensitivity due to addition or removal of some alternatives is also known as rank reversal phenomenon in MCDM literature.

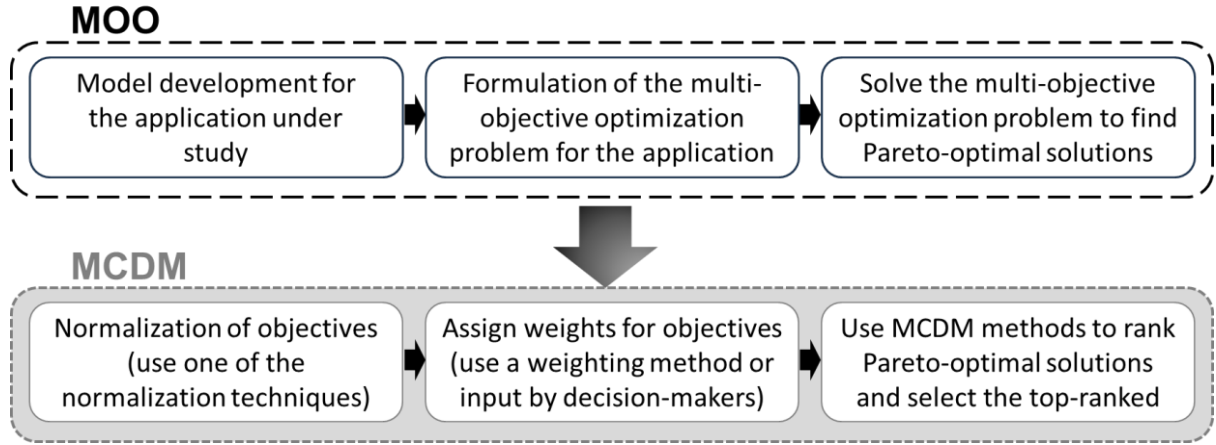

Figure 1. Main steps of MOO and MCDM

This chapter is a sequel to our recent study (Nabavi et al., 2023b), wherein we examined the sensitivity of MCDM methods to various perturbations within the decision or objective matrix (DOM). Applications covered in that study are from engineering in general. Here, we focus on the sensitivity assessment of MCDM methods for chemical engineering applications. Findings of the present study will help us to corroborate and consolidate those in Nabavi et al. (2023b). To this end, eight common or recent MCDM methods are selected to evaluate across six chemical engineering applications. The selected MCDM methods in alphabetical sequence are the combinative distance-based assessment (CODAS), complex proportional assessment (COPRAS), gray relational analysis (GRA), multi-attributive border approximation area comparison (MABAC), multi-objective optimization on the basis of ratio analysis (MOORA), preference ranking on the basis of ideal-average distance (PROBID), simple additive weighting (SAW) and technique for order of preference by similarity to ideal solution (TOPSIS).

To determine the weight of each objective within the DOM, three weighting techniques are employed in conjunction with the chosen MCDM methods. These are: criteria importance through intercriteria correlation (CRITIC), entropy, and standard deviation (StDev). This sensitivity analysis incorporates three specific types of modifications to the DOM, that is, linear transformation of an objective (LTO), reciprocal objective reformulation (ROR), and removal of alternatives (RA).

The structure of the remaining sections of this chapter is as follows: Section 2 provides an overview of the selected MCDM methods, the weighting techniques, the modifications applied to DOM, and the metrics for sensitivity analysis. Section 3 details the chemical engineering applications considered in this study. Section 4 presents results and discussion.

Finally, Section 5 concludes the chapter, summarizing the key findings and suggesting avenues for future research.

## 2. Methodologies

### 2.1. MCDM Methods

The chosen eight MCDM methods are outlined in this section.

**CODAS** method, developed by Ghorabaee et al. (2016), finds the performance score of each alternative by calculating both the Euclidean and Taxicab distances of it from the negative ideal solution. The performance score of an alternative increases with its distance from the negative ideal solution, with Euclidean distance serving as the principal metric. In instances where the performance scores of two alternatives are remarkably similar, the Taxicab distance is utilized as a secondary metric to distinguish between them. For a detailed explanation of the CODAS method, including its procedural steps and mathematical equations, refer Ghorabaee et al. (2016).

**COPRAS** method, conceived by Zavadskas and Kaklauskas (2002), employs a systematic approach to rank alternatives based on their relative importance. In this, the first step involves normalizing the original DOM using sum normalization. It is followed by multiplying each objective value by its corresponding weight to produce a weighted normalized DOM. Subsequently, the method calculates the sums of these weighted normalized values for objectives under maximization and minimization categories separately. The two sums are then utilized to ascertain the relative importance of each alternative according to specific equations of the method. See Zavadskas and Kaklauskas (2002) for details on the COPRAS method, including its procedural steps and mathematical equations.

**GRA** method, grounded in grey system theory (Deng, 1982), has several variants documented in the MCDM literature. The specific variant we reference, as outlined by Martinez-Morales et al. (2010), is notable for its independence from user inputs (e.g., weight for each objective). This particular GRA has three fundamental steps: the normalization of the DOM using max-min technique, the definition of an ideal alternative, and the calculation of the gray relational coefficient (GRC), which effectively measures the similarity between each alternative and the ideal alternative. The alternative having the highest GRC is recommended. For details on this variant of GRA, including its procedural steps and mathematical equations, see Martinez-Morales et al. (2010).

**MABAC** method, introduced by Pamučar and Ćirović (2015), centers on calculating the distance of each alternative from the border approximation area, which is derived from the product of the weighted normalized values of each objective. Like the other MCDM methods, it begins with the construction of a weighted normalized DOM from the original DOM. This is followed by the determination of the border approximation area for each objective. Finally, the distances between each alternative and the border approximation areas are computed. The alternative furthest from these areas is identified as the leading optimal solution. For an in-depth understanding of the MABAC method, including its procedural steps and mathematical formulations, refer to the work of Pamučar and Ćirović (2015).

**MOORA** method, proposed by Brauers and Zavadskas (2006), is extensively applied across various fields, including chemical engineering, to address MCDM problems. It begins with constructing a weighted normalized DOM from the original DOM. Then, the performance score for each alternative is calculated by deducting the aggregate of objectives to be minimized (i.e., cost objectives) from the sum of objectives to be maximized (i.e., benefit objectives). The alternative having the highest performance score is deemed the recommended solution. For a comprehensive understanding of the MOORA method, including its procedural steps and mathematical equations, refer Brauers and Zavadskas (2006).

**PROBID** method, proposed by Wang et al. (2021), distinguishes itself by assessing alternatives against a spectrum of ideal solutions, alongside consideration of the mean solution. This spectrum includes the most positive ideal solution followed by subsequent tiers ($2^{nd}$ most positive ideal solution, $3^{rd}$ most positive ideal solution, etc.), till the most negative ideal solution. The PROBID method involves calculating the distances between each alternative and these ideal solutions, and integrating these distances with the distance from the mean solution to generate the overall performance score. For understanding this method, including its procedural steps and mathematical equations, see Wang et al. (2021).

**SAW** method, pioneered by Fishburn (1967) and MacCrimmon (1968), stands as one of the most straightforward MCDM methods. It involves converting the original DOM into a weighted normalized version, typically facilitated by max normalization. Following this, the method sums up the weighted objective values for each alternative. The one with the highest sum (i.e., performance score) is the recommended choice by the SAW method. Further details including steps and equations of this method, are available in Wang and Rangaiah (2017).

**TOPSIS** method is one of the most widely used MCDM methods, applied across numerous disciplines. Formulated by Hwang and Yoon (1981), this method recommends the alternative, which simultaneously has the greatest Euclidean distance from the negative ideal solution (defined as the aggregation of the worst value for each objective) and the shortest Euclidean distance to the positive ideal solution (the aggregation of the best value for each objective). For details of the TOPSIS method, including its procedural steps, relevant equations, and visual illustrations, refer to the work by Hwang and Yoon (1981).

*2.2. Weighting Methods*

In all, three weighting methods are employed in this work.

**CRITIC** weighting method considers the pairwise correlation between objectives within the DOM along with the standard deviation of each objective, to derive weights. Its underlying principle is to capture both the diversity of objective values and their interdependencies, thereby offering a more comprehensive approach to weight assignment. For a detailed explanation of its rationale, steps, and equations, refer to Diakoulaki et al. (1995).

**Entropy** weighting method is grounded in the probabilistic approach to quantify informational uncertainty. It assigns a higher weight to an objective having a larger variation in its values across the DOM. As highlighted in the review by Hafezalkotob et al. (2019), entropy method is frequently cited as the most popular weighting technique in MCDM. Hwang and Yoon (1981) presents the procedural steps, equations, and illustrations of this method.

**StDev** weighting method is an intuitive approach for determining the weights of objectives in MCDM. It calculates the standard deviation of values for each objective in DOM, to capture the variability or dispersion of values around the mean. The StDev method assigns greater weight to an objective with a higher standard deviation. See Wang et al. (2020) for the steps and equations of the method.

*2.3. DOM Modifications*

The 3 modifications: LTO, ROR and RA considered for sensitivity analysis of MCDM methods are outlined in this section. One modification at a time is considered.

**LTO**: Pamučar and Ćirović (2015) utilized the concept of the independence of the value scale (IVS) for analyzing the consistency of MCDM methods. IVS suggests that the rankings produced by an MCDM method should remain unchanged when the units of measurement (or scales) of the objectives undergo linear transformations. In this chapter, an LTO in the form of $y = Ax + B$ is applied, where A and B represent positive constants, and x and y denote the original and transformed values of the objectives in DOM, respectively.

**ROR**: When an objective is equivalently reformulated (e.g., converting maximization of profit to minimization of 1/profit), the ranking of alternatives by an MCDM method should remain unchanged. Essentially, the formulation of objectives, whether as smaller-the-better or larger-the-better, ought not to alter the rankings. Pamučar and Ćirović (2015) discussed choosing a forklift in a logistic center, where one objective is the speed of lifting loads (S) measured in *meters per minute* (larger-the-better). However, this can be converted to a smaller-the-better objective using 1/S, measured in *minutes per meter,* signifying the time needed to lift a load by one meter. In this chapter, a ROR transformation, y = 1/(x + A), is applied, where A is a constant (possibly zero), and x and y denote the original and transformed values of the objectives in DOM, respectively.

**RA**: As mentioned in Section 1, in a rational decision-making process, when some solutions are removed from the DOM, the relative rankings of the remaining solutions should not change. However, this modification poses a significant challenge to many MCDM methods, as their calculations often rely on the positive and/or negative ideal solutions, which could include portion of the solutions that have been removed from the DOM. To assess this sensitivity, this chapter details an experiment where 10% of the solutions from the DOM are randomly removed.

## 2.4. Sensitivity Analysis Metrics

To measure how the weights by a weighting method change in response to a modification in DOM, the average absolute fractional deviation (AAFD) for weights, defined in Equation (1) below, is employed:

$$AAFD_w = \frac{1}{n} \sum_{j=1}^{n} \frac{|WO_j - WM_j|}{\left|\frac{WO_j + WM_j}{2}\right|} \quad (1)$$

Here, $n$ is the number of objectives, and $WO_j$ and $WM_j$ are the calculated weights for the $j^{th}$ objective in the original and modified DOMs, respectively. Minimal sensitivity requires zero or negligible value of $AAFD_w$.

Analogously, as shown in Equation (2), the AAFD for the top-ranked alternative is used to quantify the discrepancy between the top-ranked alternatives derived from the original and the modified DOMs by a MCDM method.

$$AAFD_o = \frac{1}{n}\sum_{j=1}^{n}\frac{|OO_j - OM_j|}{\left|\frac{OO_j + OM_j}{2}\right|} \tag{2}$$

Here, $OO_j$ and $OM_j$ are the values of the $j^{th}$ objective in the top-ranked alternative from the original and the modified DOMs, respectively, by a MCDM method. Minimal sensitivity requires zero or negligible value of $AAFD_o$. When needed, the $AAFD_o$ can also be applied to alternatives ranked second, third, and so forth.

Spearman's rank correlation coefficient ($r_s$), computed using Equation (3), is utilized to measure the difference between the ranking of all alternatives found by a MCDM method using the original and the modified DOMs. Subsequently, with Equation (4), a t-test is performed to calculate the p-value for $r_s$. Generally, a $r_s$ is considered statistically significant if its p-value is less than 0.05, providing robust evidence to support its acceptance.

$$r_s = \frac{\sum_{i=1}^{m}(x_i - \bar{x})(y_i - \bar{y})}{\sqrt{\sum_{i=1}^{m}(x_i - \bar{x})^2 \sum_{i=1}^{n}(y_i - \bar{y})^2}} \tag{3}$$

$$t = \frac{r_s\sqrt{m-2}}{\sqrt{1-r_s^2}} \tag{4}$$

Here, $m$ is the number of all alternatives; $x_i$ and $y_i$ are the ranks of each alternative before and after the DOM modification; and $\bar{x}$ and $\bar{y}$ are the mean ranks of these alternatives. A perfect positive correlation is indicated by a $r_s = 1$.

## 3. Applications

The 6 applications of MCDM in chemical engineering, employed in this work, are briefly described in this section.

***Polyethylene Waste Co-gasification:*** The waste gasification involves the decomposition of a solid waste in the presence of a gasifying agent (air, steam) with heating. Its main products are hydrogen and carbon dioxide. Due to the high content of carbon and hydrogen, waste plastics are one of the desirable feed stocks for this process. The use of these feeds is important from both environmental and economic points of view. Gasification is a thermochemical process, and its efficiency and performance are affected by various variables. This dataset used in this study is for the co-gasification of low- and high-density polyethylene waste with different composition ratios. The MCDM problem involved 11 alternatives and 5 energy, environmental and economic criteria of gasification performance; the indicators include hydrogen production rate, energy efficiency, carbon dioxide emission, heating value of syngas and purchasing cost (Hasanzadeh et al., 2023). In the LTO, energy efficiency is modified by y = 2x + 3; and in the ROR, cost is transformed by y = 1/x from minimization (min) to maximization (max). The alternative L20H8 (20% wt low density and 80% wt high density polyethene) is removed in case of RA.

*Synthesis Gas Production from Biomass:* In the future, supplying energy from renewable sources will be one of the main solutions to meet the needs of industries and address environmental concerns. The use of biomass as a feedstock for energy production processes has been studied for decades. Converting gasification of biomass into valuable intermediates such as synthesis gas will create added value. Synthesis gas is rich in hydrogen and carbon monoxide; in addition to being used in the combustion process to provide thermal energy, it can be used as a feed for the synthesis of basic petrochemicals. Choosing the type of biomass for the synthesis gas production is important due to the variation in the composition of biomass. In this dataset for MCDM application, the gas composition and efficiency of the system are investigated for different biomasses as gasification fuels. The original data involves 13 alternative biomasses; of these, only 7 are non-dominated and are used in the present study. There are 7 criteria; of them, the benefit criteria are hydrogen and methane content in the synthesis gas, hydrogen yield, cold gas efficiency, and exergy efficiency. Carbon monoxide and dioxide in the synthesis gas are cost criteria (Mojaver et al., 2019). Hydrogen and carbon monoxide content is modified by y = 2x + 3 in LTO; while carbon dioxide content and cold gas efficiency are transformed by y = 1/x from min to max and max to min, respectively. To study RA, Legume straw biomass is removed from the dataset.

*Thermal Cracking Reactor for Olefins:* The thermal cracking process has been the only industrial process for the production of olefins on a large scale. It is a fundamental and upstream unit in the petrochemical industry. The thermal cracking reactor is the most complex part of this plant. MOO of thermal cracking reactor with liquefied petroleum gas (LPG) feed was studied by Nabavi et al. (2009; 2011) considering simultaneous maximization of production of ethylene and propylene and minimization of heat duty. The dataset used in the present study is from Nabavi et al. (2023a); it contains 60 non-dominant solutions with two benefit criteria and one cost criterion. In this dataset, ethylene production is modified by y = 2x + 3 for LTO. Propylene production is transformed from max to min by y = 1/x in ROR. In case of RA, six alternatives are removed from 60 of total non-dominated solutions.

*Cumene Production Process:* Cumene is an important petrochemical intermediate to produce phenol and acetone, which account for about 98% of cumene produced in the world. It is produced from the reaction of benzene and propylene in gas phase in catalytic fixed bed reactors. Cumene production process has been optimized for both operation and design, considering operational, economic, and environmental objectives (Sharma, et al., 2013; Flegiel, et al., 2015). The dataset of non-dominant optimal solutions from Amooey et al. (2023) for cumene production process is chosen for the MCDM study. It has 120 alternatives and three environmental and economic criteria, namely, damage index, material loss and total capital cost that should be minimized. In this data set, total capital cost was modified by $y = 2x + 3$ in LTO and material loss is transformed from min to max by $y = 1/x$ in ROR. 12 solutions are removed in the case of RA.

*Material for Cryogenic Tank Design:* Cryogenic tanks are used for storage and transportation of gases like nitrogen, oxygen, hydrogen, helium and argon. These tanks should be made of special martials to be effective. The number of available materials, together with the complex relationships between the various selection parameters, makes it difficult to select the most appropriate material. The materials should have good weldability and processability, lower density and specific heat, smaller thermal expansion coefficient and thermal conductivity, and adequate toughness at the operating temperature (Dehghan-Manshadi, et al., 2007). The dataset in Mousavi-Nasab et al. (2018) is for the material selection for liquid nitrogen tank. There are 7 alternatives, each with 7 criteria; of these, toughness index, yield strength and Young's modulus are benefit type whereas density, thermal expansion index, thermal Conductivity and specific heat are cost type. In this case, Young's modulus and density is modified by $y = 2x + 3$ in LTO. Next, yield strength is transformed from max to min and thermal conductivity is modified from min to max by $y = 1/x$ in ROR. Alternative Al5052-0 is removed from the DOM to study effect of RA.

*Material for High Temperature Oxygen-Rich Environment:* This MCDM application is on selection of a suitable material for an equipment, which needs to be designed for operation in a high-temperature oxygen-rich environment. The material selection matrix involved 6 alternatives and 4 criteria (Rao, 2006); among these, hardness, machinability rating and corrosion resistance should be maximized, and the cost should be minimized (Mousavi-Nasab et al., 2018). In this case, machinability rating is converted by $y = 2x + 3$ in LTO. While hardness is transformed from max to min by $y = 1/x$. Material two is removed from DOM in the case of RA.

## 4. Results and Discussion

Effect of the three modifications in DOM on MCDM and weighting methods is discussed in the following sub-sections: effect on weights, effect on the top-ranked (i.e., rank 1) alternative, effect on top 3 (i.e., rank1, 2, and 3) alternatives, effect on ranking of all alternatives, and summary of effects.

### *4.1. Effect of Modifications on Weighting Methods*

To analyze the effect of the three modifications in DOM on the weights calculated by the entropy, CRITIC and StDev methods, mean $AAFD_w$ between the weights of criteria based on the original DOM and the weights of criteria based on each of the modified DOMs are computed. As shown in Figure 2, weights calculated by the StDev and CRITIC methods for all three modified DOMs are more similar to each other compared to those found by the entropy method. In the case of LTO, the mean $AAFD_w$ is zero for both StDev and CRITIC methods. In other words, the same weights are obtained for both the original and modified DOM, and the calculated weights are unaffected by LTO. In the case of ROR and RA, the mean $AAFD_w$ is not zero; in these two modifications of DOM, the weights calculated by the StDev and CRITIC methods are different from those values based on the original DOM. The trend of mean $AAFD_w$ is different for the weights calculated by the entropy method. All three modifications of DOM affect the calculated weights for the criteria and the effect of ROR is greater than the effect of LTO and RA. In summary, the weights calculated by the entropy method are more sensitive to LTO, ROR and RA modifications compared to StDev and CRITIC methods, which is consistent with the finding of Nabavi et al. (2023b).

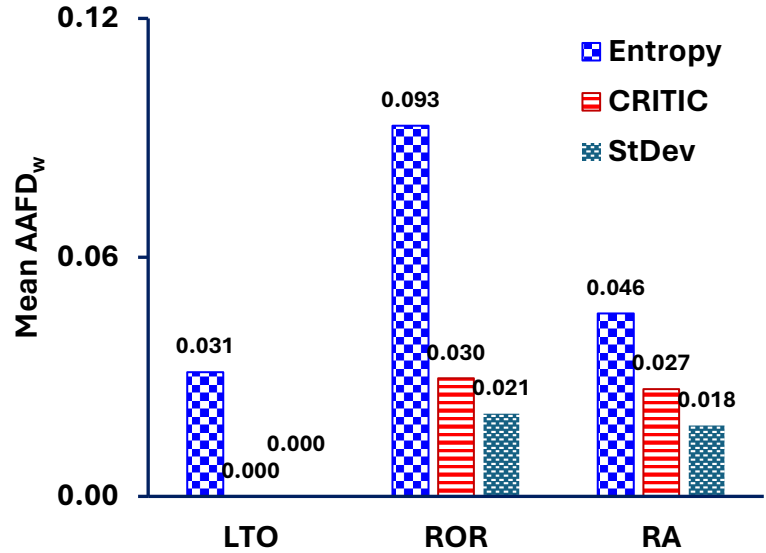

Figure 2. Effect (quantified by mean $AAFD_w$) of LTO, ROR, and RA on the weights by entropy, CRITIC and StDev methods,

## 4.2. Effect of Modifications on Rank 1 Alternative

Figure 3 shows the effect of LTO, ROR, and RA on the rank 1 alternative by different MCDM methods with entropy, CRITIC and StDev weights. Frequency in plots A, C and E, is the number of changes in the rank 1 alternative due to LTO, ROR, or RA in the 6 applications. The maximum frequency value is found to be 3, which means rank 1 alternative by the combined weighting and MCDM method employed is affected by the modification in 3 out of 6 applications tested. Frequency will be zero if there is no change in the rank 1 alternative; then, $AAFD_o$ in each application and its sum will also be zero. Low/zero frequency and low/zero mean $AAFD_o$ are desirable for a less sensitive (i.e., more robust) MCDM method. GRA is not affected by weight changes because the variant of GRA used in this study does not require weights. LTO has the least effect on the rank 1 alternative identified by all 8 MCDM methods with all weighting methods; RA has slightly more effect; whereas ROR has the most effect.

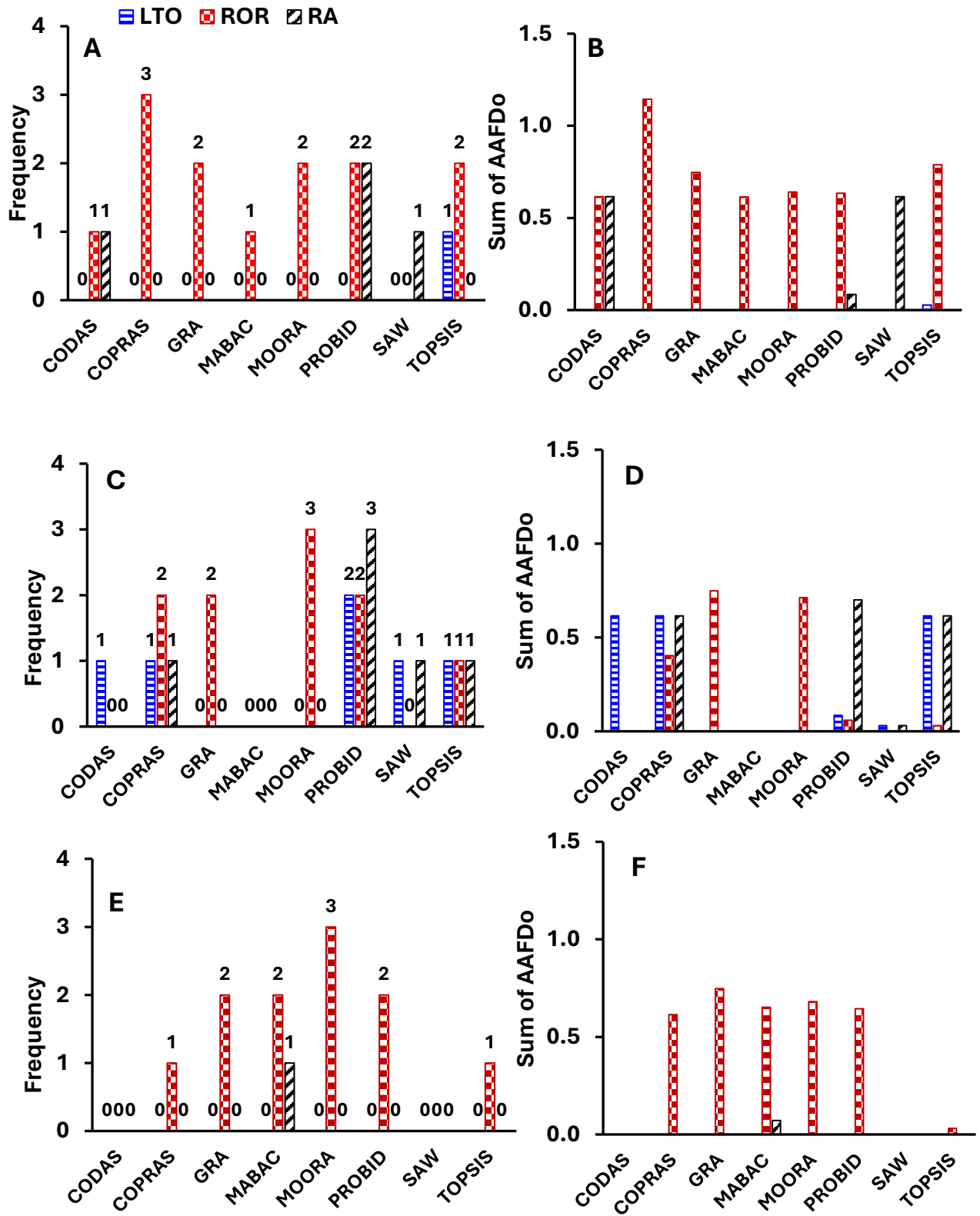

Figure 3. Effect of LTO, ROR and RA on the rank 1 alternative by 8 MCDM methods using the weights by entropy (plots A and B), CRITIC (plots C and D) and StDev (plots E and F)

methods. In plots A, C and E, y-axis is frequency whereas it is sum of $AAFD_o$ in plots B, D and F.

Table 1. Statistics on the effect of LTO, ROR, and RA on the Rank 1 alternative by all 8 MCDM Methods with entropy, CRITIC and StDev weights.

| Weights | Modification | Frequency | Sum of $AAFD_0$ |
|---|---|---|---|
| Entropy | LTO | 1 | 0.03 |
|  | ROR | 13 | 5.18 |
|  | RA | 3 | 1.32 |
|  | **Total** | **17** | **6.53** |
| CRITIC | LTO | 6 | 1.96 |
|  | ROR | 10 | 1.95 |
|  | RA | 6 | 1.96 |
|  | **Total** | **26** | **5.88** |
| StDev | LTO | 0 | 0 |
|  | ROR | 11 | 3.37 |
|  | RA | 1 | 0.07 |
|  | **Total** | **12** | **3.44** |

Statistics in Table 1 show that the LTO and RA have no or little effect when StDev weights are used with the MCDM methods. However, the effect of ROR is greater when using entropy and StDev weights. Interestingly, the sum of $AAFD_o$ is almost same for LTO, ROR and RA when CRITIC weights are used with the MCDM methods. Overall, changes in the rank 1 alternative by the 8 methods are less with StDev weights compared to those with CRITIC and entropy weights. Considering both the frequency and $AAFD_o$ results for finding rank 1 alternative, less sensitive methods are GRA (without any weights); MABAC, CODAS, and SAW when using entropy and CRITIC weights; CODAS, SAW, and TOPSIS when using StDev weights (Figure 3). Comparison of total frequency and sum of $AAFD_o$ values in Table 1 shows that the effect of modifications on the top alternative by all 8 MCDM methods is less (i.e., more similar alternatives are selected) when the StDev weights are used compared to when the entropy and CRITIC weights are used.

*4.3. Effect of Modifications on Top 3 Alternatives*

Next, the effect of the modifications on the top 3 alternatives by each of the 8 MCDM methods with entropy, CRITIC and StDev weights was investigated. Figure 4 shows this effect in terms of total frequency of changes (i.e., sum for top 3 alternatives) and total $AAFD_o$ (i.e., sum for top 3 alternatives). As before, ranking by GRA is unaffected by weights and hence its results for total frequency and total $AAFD_o$ in the top, center and bottom plots in Figure 4 are identical. Considering both the frequency and $AAFD_o$ results for finding top 3 alternatives, the less sensitive methods with the entropy and CRITIC weights are MABAC and SAW; with StDev weights, they are SAW, CODAS and COPRAS.

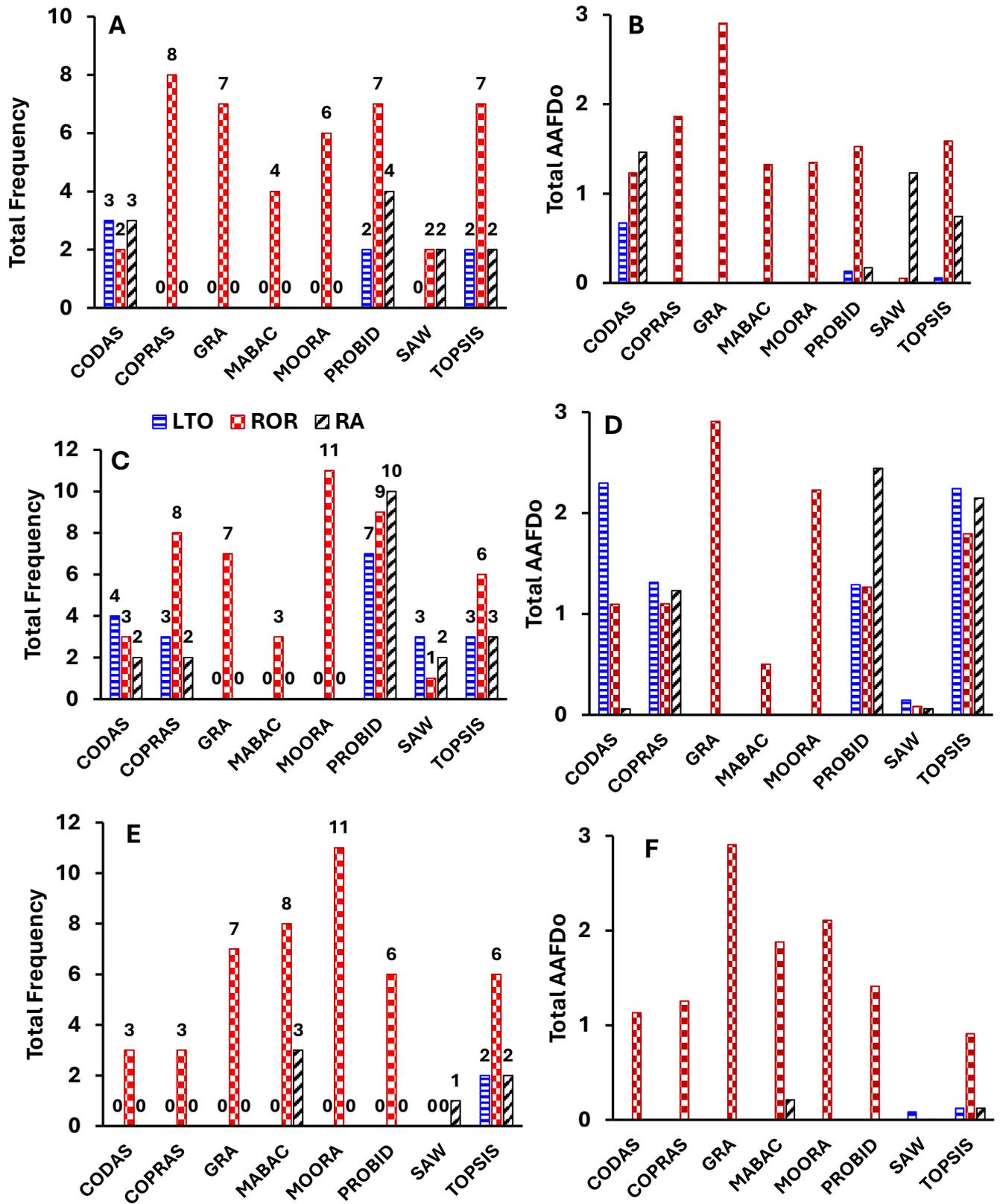

Figure 4. Effect of LTO, ROR and RA on the top 3 alternatives by 8 MCDM methods using weights by the entropy method (plots A and B), CRITIC (plots C and D) and StDev (plots E

and F). In plots A, C and E, y-axis is total frequency whereas it is total $AAFD_o$ in plots B, D and F.

Statistics in Table 2 clearly show the following effects on the top 3 alternatives identified by MCDM methods under LTO ROR, and RA modifications: (a) LTO has the least effect in case entropy and StDev weights, (b) compared to both LTO and RA, ROR has more effect irrespective of which of the three weighting methods is used, and (c) overall, effect of LTO, ROR and RA is more if CRITIC weights are employed.

Table 2. Statistics on the effect of LTO, ROR and RA on top 3 alternatives by all 8 MCDM methods with entropy, CRITIC and StDev weights.

| Weights | Modification | Frequency | Sum of $AAFD_0$ |
|---|---|---|---|
| Entropy | LTO | 7 | 0.86 |
|  | ROR | 43 | 11.83 |
|  | RA | 11 | 3.61 |
|  | **Total** | **61** | **16.29** |
| CRITIC | LTO | 20 | 7.29 |
|  | ROR | 48 | 10.98 |
|  | RA | 19 | 5.94 |
|  | **Total** | **78** | **24.21** |
| StDev | LTO | 2 | 0.21 |
|  | ROR | 44 | 11.60 |
|  | RA | 6 | 0.34 |
|  | **Total** | **52** | **12.15** |

*4.4. Effect of Modifications on Ranking of All Alternatives*

The effect of LTO, ROR, and RA on ranking of all alternatives is quantified by Spearman's rank correlation coefficient between the ranking of the original dataset and that of the modified dataset. Table 3 summarizes the average correlation coefficient and p-value for 6 applications. The Spearman's rank correlation coefficient of +1 indicates perfect positive relationship; this implies there is no change in the ranking of all alternatives due to the modification in the dataset. Further, Spearman's rank correlation coefficient with p-value less than 0.05 is said to be statistically significant, indicating strong evidence that the calculated correlation coefficient can be accepted.

The average Spearman's rank correlation coefficient and p-value for all the 8 MCDM methods are given in the last row of Table 3. Spearman's rank correlation coefficient is greater than 0.98 in the case of LTO and RA for all three weighting methods, whereas it is 0.95 to 0.97 in the case of ROR. These values indicate that ROR has more effect on ranking of all alternatives, regardless of the weighting methods. This observation coincides with the results in Figure 3 and Table 1 on the top-ranked alternative, as well as Figure 4 and Table 2 on top 3 alternatives, where ROR has the largest effect, irrespective of weighting method used.

*4.5. Summary of Effects of Modifications*

The results of this study demonstrate that weights calculated by the entropy method are more sensitive to LTO, ROR, and RA, compared to those by the StDev and CRITIC method, for all 6 applications tested (Figure 2). Additionally, it is found that ROR has the greatest effect on the ranking of alternatives by MCDM methods (Tables 1, 2, and 3). Lastly, comprehensively considering the frequency and $AAFD_o$ values from Figures 2 and 3, and Spearman's rank correlation coefficients and p-values from Table 3, GRA (without any weights), MABAC, CODAS, and SAW with entropy or CRITIC weights, and CODAS, SAW, and TOPSIS with StDev weight are shown to be less sensitive to the modifications in DOM. These findings are generally consistent with those reported in Nabavi et al. (2023b).

Table 3. Effect of LTO, ROR and RA on Ranking of All Alternatives: Average of Spearman's Rank Correlation Coefficient (with p-Value in brackets)

| MCDM Method | Entropy Weights | | | | CRITIC Weights | | | | StDev Weights | | | |
|---|---|---|---|---|---|---|---|---|---|---|---|---|
| | LTO | ROR | RA | Average | LTO | ROR | RA | Average | LTO | ROR | RA | Average |
| CODAS | 0.9841 (0.00088) | 0.9850 (0.00042) | 0.9903 (0.00080) | **0.9865 (0.00070)** | 0.9333 (0.01997) | 0.9363 (0.01197) | 0.9813 (0.00623) | **0.9503 (0.01272)** | 0.9986 (4.7E-40) | 0.9834 (7.6E-05) | 0.9891 (0.00080) | **0.9904 (0.00029)** |
| COPRAS | 0.9994 (4.7E-40) | 0.9811 (0.00088) | 0.9999 (2.3E-25) | **0.9935 (0.00029)** | 0.9814 (0.00050) | 0.9570 (0.00700) | 0.9802 (0.00314) | **0.9729 (0.00355)** | 0.9993 (4.7E-40) | 0.9745 (0.00088) | 0.9995 (2.3E-25) | **0.9911 (0.00029)** |
| GRA | 1.0000 (4.7E-40) | 0.9691 (0.00122) | 1.0000 (2.3E-25) | **0.9897 (0.00041)** | 1.0000 (4.7E-40) | 0.9691 (0.00122) | 1.0000 (2.3E-25) | **0.9897 (0.00041)** | 1.0000 (4.7E-40) | 0.9691 (0.00122) | 1.0000 (2.3E-25) | **0.9897 (0.00041)** |
| MABAC | 0.9994 (4.7E-40) | 0.9710 (0.00228) | 0.9995 (2.3E-25) | **0.9900 (0.00076)** | 1.0000 (4.7E-40) | 0.9202 (0.00703) | 0.9967 (1.5E-8) | **0.9723 (0.00234)** | 1.0000 (4.7E-40) | 0.9543 (0.00708) | 0.9995 (2.3E-25) | **0.9846 (0.00236)** |
| MOORA | 0.9997 (4.7E-40) | 0.9744 (0.00228) | 0.9999 (2.3E-25) | **0.9913 (0.00076)** | 0.9818 (0.00114) | 0.9684 (0.00130) | 0.9901 (0.00080) | **0.9801 (0.00108)** | 0.9997 (4.7E-40) | 0.9590 (0.00428) | 0.9997 (2.3E-25) | **0.9861 (0.00143)** |
| PROBID | 0.9997 (4.7E-40) | 0.9352 (0.00957) | 0.9808 (0.00160) | **0.9719 (0.00373)** | 0.9752 (0.00121) | 0.9422 (0.00814) | 0.9680 (0.00394) | **0.9618 (0.00443)** | 0.9932 (7.6E-5) | 0.9776 (0.00114) | 0.9991 (2.3E-25) | **0.9900 (0.00040)** |
| SAW | 0.9994 (4.7E-40) | 0.9988 (4.7E-40) | 0.9903 (0.00080) | **0.9962 (0.00027)** | 0.9936 (7.6E-5) | 0.9931 (7.6E-5) | 0.9993 (2.3E-25) | **0.9953 (5.1E-5)** | 0.9996 (4.7E-40) | 0.9991 (4.7E-40) | 0.9997 (2.3E-25) | **0.9995 (7.8E-26)** |
| TOPSIS | 0.9997 (4.7E-40) | 0.9608 (0.00308) | 0.9904 (0.00080) | **0.9836 (0.00130)** | 0.9754 (0.00228) | 0.9388 (0.00957) | 0.9611 (0.01207) | **0.9584 (0.00797)** | 0.9991 (4.7E-40) | 0.9835 (0.00042) | 0.9993 (2.3E-25) | **0.9940 (0.00014)** |
| **Average R (p-value)** | **0.9977 (0.00011)** | **0.9719 (0.00247)** | **0.9939 (0.00050)** | | **0.9801 (0.00315)** | **0.9531 (0.00579)** | **0.9846 (0.00327)** | | **0.9987 (9.5E-6)** | **0.9751 (0.00189)** | **0.9982 (0.00010)** | |

## 5. Conclusions

In this chapter, we investigated the impact of three types of DOM modifications—LTO, ROR, and RA—on the rankings of alternatives, focusing on both the top-ranked and the top 3 alternatives, along with the overall rankings. We utilized eight common or recent MCDM methods that are integrated with entropy, CRITIC, or StDev weighting methods across six chemical engineering applications. Our findings revealed that: First, the weights generated using entropy method demonstrated higher sensitivity to LTO, ROR, and RA modifications when compared to those derived from the StDev and CRITIC methods. Second, the analysis identified ROR as having the most significant influence on the ranking of alternatives by MCDM methods. Last but not least, GRA (without any weights), MABAC, CODAS, and SAW with entropy or CRITIC weights, and CODAS, SAW, and TOPSIS with StDev weight, exhibit lower sensitivity, thereby indicating better robustness, to modifications in DOM.